%% file: Hgformer.tex
\documentclass[conference]{IEEEtran}
\IEEEoverridecommandlockouts
\usepackage{amsmath,amssymb,amsfonts}
\usepackage{algorithmic}
\usepackage{textcomp}
\usepackage{microtype}
\usepackage{graphicx}
\usepackage{subfigure}
\usepackage{balance}
\usepackage{booktabs} 
\usepackage{amsthm}
\usepackage{array}
\usepackage{graphicx}
\usepackage{clrscode}
\usepackage{subfigure}
\usepackage{multirow}
\usepackage{multicol}
\usepackage{float}
\usepackage{color}
\usepackage{xcolor}
\usepackage{amsopn}
\usepackage{mathrsfs}
\usepackage{mathtools}
\usepackage{amsmath}
\usepackage{booktabs}
\usepackage{arydshln}
\usepackage{hyperref}
\usepackage{blkarray}
\usepackage{enumerate}
\usepackage{courier}
\usepackage{mathrsfs}
\usepackage{rotating}
\usepackage{bm}
\usepackage{subfigure}
\usepackage{array}
\usepackage{ragged2e}
\usepackage{hyperref}
\usepackage{amsmath}
\usepackage{threeparttable}

\definecolor{hengpink}{cmyk}{0, 0.7808, 0.4429, 0.1412}

\def\BibTeX{{\rm B\kern-.05em{\sc i\kern-.025em b}\kern-.08em
    T\kern-.1667em\lower.7ex\hbox{E}\kern-.125emX}}
\begin{document}
\title{Hgformer: Hyperbolic Graph Transformer for Recommendation}

\author{
\IEEEauthorblockN{
Xin Yang$^{1}$, Xingrun Li$^{2}$, Heng Chang$^{3}$, Jinze Yang$^{4}$, Xihong Yang$^{5}$, Shengyu Tao$^{3}$, Ningkang Chang$^{2}$,\\
Maiko Shigeno$^{1}$,  Junfeng Wang$^{5}$, Dawei Yin$^{5}$, Erxue Min$^{5\dagger}$}
\IEEEauthorblockA{
$^1$University of Tsukuba, Japan, $^2$Kyoto University, Japan, $^3$Tsinghua University, China, \\
$^4$The University of Tokyo, Japan, $^5$Baidu Inc., China \\
\{s2330128@u.tsukuba.ac.jp, li.xingrun.65h@st.kyoto-u.ac.jp, changh17@tsinghua.org.cn, \\
tsy22@mails.tsinghua.edu.cn, yangjinze@g.ecc.u-tokyo.ac.jp, nk.chang@amp.i.kyoto-u.ac.jp, \\
maiko@sk.tsukuba.ac.jp, wangjunfeng@baidu.com, \\
yindawei@acm.org, erxue.min@gmail.com\}} 
\thanks{$^\dagger$ Corresponding author: Erxue Min (erxue.min@gmail.com).}
}

\maketitle

\input{sections/abstract}

\input{sections/introduction}

\input{sections/preliminaries}

\input{sections/Hgformer}

\input{sections/experiments}
\input{sections/relatedworks}
\section{Proof of Theorems}\label{proof}
\subsection{Proof of Theorem~4.1}
\input{proofs/theorem1proof}
\subsection{Proof of Lemma~4.2}

\input{proofs/lemma2proof}
\subsection{Proof of Theorem~4.2}
\input{proofs/theorem2proof}

\section{Conclusion}
In this research, we introduce the hyperbolic manifold into transfer learning and propose a cross-domain recommendation model using BiTGCF as the base model. Additionally, for better performance in graph collaborative filtering and numerical stability, we design a new simplified GCN with dense connections in the propagation layer. Our transfer learning structure consists of two parts: knowledge transfer between users and knowledge transfer between items. For knowledge transfer between users, we constructed a completely new transfer function. For knowledge transfer between items, which was not considered in the base model, we propose a knowledge transfer method on hyperbolic manifold by improving the transformer model. Through experiments, we identify the optimal model. The significant performance improvements across various datasets compared to the baseline models demonstrate the effectiveness of our proposed model.\\
\indent On the other hand, although we propose a knowledge transfer method based on the transformer model, the results are not as satisfactory as expected, and the reasons remain unclear. The mechanism of transformer has been proven effective in various domains. Exploring how to improve and integrate the transformer into transfer learning structure as a module is a valuable research direction for the future.


\bibliographystyle{IEEEtran}
\balance
\bibliography{Hgformer}

\end{document}

%% file: sections/abstract.tex
\begin{abstract}
The cold start problem is a challenging problem faced by most modern recommender systems. By leveraging knowledge from other domains, cross-domain recommendation can be an effective method to alleviate the cold start problem. However, the modelling distortion for long-tail data, which is widely present in recommender systems, is often overlooked in cross-domain recommendation. In this research, we propose a hyperbolic manifold based cross-domain collaborative filtering model using BiTGCF as the base model. We introduce the hyperbolic manifold and construct new propagation layer and transfer layer to address these challenges. The significant performance improvements across various datasets compared to the baseline models demonstrate the effectiveness of our proposed model.
\end{abstract}
\begin{IEEEkeywords}
Collaborative Filtering, Hyperbolic Learning, Graph Neural Networks
\end{IEEEkeywords}

%% file: sections/introduction.tex



\section{Introduction}
Recommender systems have become an indispensable part of our daily life, serving as fundamental tools for personalized information filtering and prioritization \cite{w&d,youtube,taobao}. The core of a recommender system is to predict whether a user will engage with an item, such as by clicking, rating, or purchasing it. In this context, Collaborative Filtering (CF) \cite{rendle2012bpr,xue2017deep,zheng2018spectral,lightgcn}, which leverages past interactions between users and items to make these predictions, remains an essential component to deliver effective personalized recommendations.
The interaction patterns between users and items in CF tasks naturally form a graph structure, motivating researchers to investigate the use of Graph Neural Networks (GNNs) \cite{lightgcn,ngcf,hgcf}, which has proven significant advantages in modeling graph structures\cite{kipf2016semi,hamilton2017inductive}. 

\begin{figure}[!tb]
    \centering
 \includegraphics[width=0.5\textwidth]{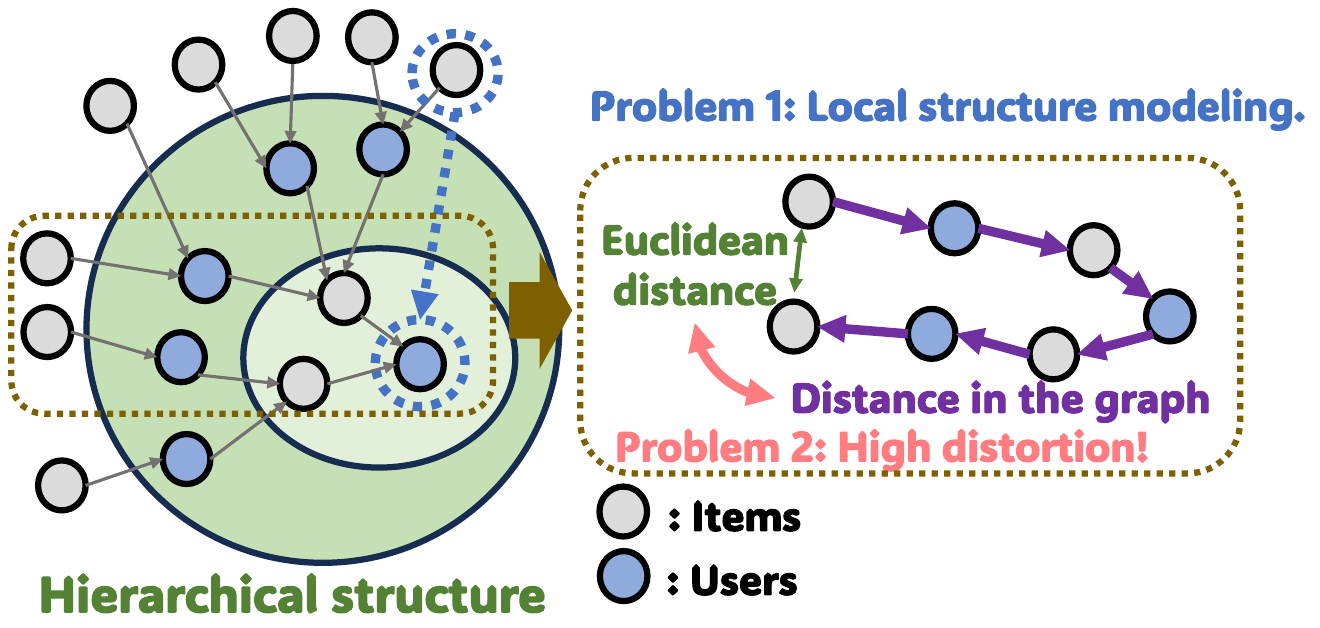}
    \caption{This figure highlights two key challenges that traditional GNNs often face when dealing with long-tail items. The first one is the message-passing paradigm in local neighborhoods, which prevents the model from transferring information from less popular items to most users, leading to poor recommendations for these items. The second one is that the embeddings in Euclidean space may fail to capture the long distance in graphs, and will lead to serious information distortion in graph structure.}
    \label{fig:motivation}
\end{figure}

Despite the significant attention and fruitful outcomes in this field, most existing GNNs normally assume that the degree of each node is balanced. However, data in the realm of recommender systems generally exhibit a long-tail distribution \cite{park2008long,park2012adaptive,yin2012challenging,zhao2023alleviating}: a small portion of items are highly popular with numerous users, whereas most other items attract relatively few users. 
Recent studies have also shown that GNNs-based methods perform well in recommending popular items (head items), but often struggle to perform as effectively with less popular items (tail items) \cite{hgcf,hicf}.
This issue primarily arises from two factors, which are described in Fig.~\ref{fig:motivation}:
\textbf{i) Local structure modeling:} GNN-based models normally follow the neighborhood aggregation scheme and tend to be biased towards nodes with high degree \cite{liu2023generalized}. More precisely, in the task of CF, head items that are interacted with by many users typically have higher-quality representations, while the vast majority of tail nodes with few interactions are likely to be underrepresented \cite{longtailexpert}. Several GNN frameworks have been proposed to mitigate these degree biases by introducing designated architectures or training strategies specifically for low-degree nodes \cite{hgcf,longtailexpert,zhao2023alleviating}, but they still fail to capture the global information as transformers \cite{graphformer,difformer,sgformer,nodeformer} and thus achieve sub-optimal results. 
\textbf{ii) Embedding distortion:} Most existing methods encode items and users into Euclidean space \cite{ngcf,lightgcn}, which is a flat geometry with a polynomial expanding capacity. Data with the long-tail distribution can be traced back to hierarchical structures \cite{hie}, whose number of neighbors increases exponentially. As a result, encoding these data via Euclidean space naturally incurs information loss and could subsequently deteriorate the performance of downstream tasks.
In contrast, hyperbolic manifold, a non-Euclidean space characterized by constant negative curvature, allows the space to expand exponentially with the radius, making it particularly well-suited for representing tree-like or hierarchical structures \cite{hgcn,hnn}.
For this reason, in recent years, significant advances have been made in hyperbolic neural networks to better handle the problem of long-tail distribution \cite{hgcn,hgcc,hcl}.

So far, most research in recommender systems has focused on addressing either one of the two problems but has not managed to tackle both simultaneously. Inspired by the successful application of Graph Transformers in graph and node classification tasks \cite{graphformer,nodeformer,sgformer}, this study proposes a novel Graph Transformer architecture in the hyperbolic manifold for Collaborative Filtering. Although the idea of extending Graph Transformers to hyperbolic space for recommendations is intriguing, it poses several challenges that must be overcome:


\textbf{$\bullet$\;}\textbf{No well-defined parameter-free graph convolution in hyperbolic space.} Parameter-free message-passing paradigms such as LightGCN \cite{lightgcn} have shown superior advantages in CF, however, most existing hyperbolic variants of LightGCN, such as HGCF, HRCF, and HICF \cite{hgcf,hrcf,hicf} require to first project embeddings in the hyperbolic manifold back to the tangent space for subsequent graph convolution, which causes information loss and limits their performance. 

\textbf{$\bullet$\;}\textbf{No well-defined hyperbolic self-attention mechanism for collaborative filtering.} 
Although there are currently some definitions of hyperbolic attention \cite{hgat,yang2024hypformer}, in the field of Collaborative Filtering, the user-item interaction structure is represented as a bipartite graph, existing methodologies are inadequate, and hyperbolic self-attention mechanism tailored for CF tasks has not yet been investigated.


\textbf{$\bullet$\;}\textbf{ Scalability issue of hyperbolic self-attention mechanism.} 
In recommender systems, real-world graphs are often large-scale, which poses a significant challenge in efficiency when applying transformer architectures with quadratic time complexity. Although many studies have tackled the scalability problem of graph transformers in Euclidean space \cite{nodeformer,sgformer,difformer}, the solution for linear computational complexity of hyperbolic self-attention is still under-explored. 

To solve these challenges, we propose a new \textbf{\underline{H}}yperbolic \textbf{\underline{G}}raph Trans\textbf{\underline{former}} framework called \textbf{Hgformer}. 
For the first challenge, we propose the Light Hyperbolic Graph Convolutional Network (LHGCN), which performs graph convolution entirely in the hyperbolic manifold; For the second challenge, we propose a novel hyperbolic transformer architecture tailored for CF tasks, which consists of a cross-attention layer and a hyperbolic normalization layer; For the last challenge, we propose an unbiased approximation approach to reduce the computational complexity of hyperbolic self-attention to the linear level.
Numerical experiments show that our proposed model performs better than the leading CF models and remarkably mitigates the long-tail issue in CF tasks.
We summarize our contributions as follows:
\begin{itemize}
\item We propose a novel LHGCN model to capture the local structure of each node within the hyperbolic manifold.
\item  We introduce a hyperbolic transformer architecture to facilitate global information learning.
\item We propose a linearization technique to reduce the computational complexity of hyperbolic cross-attention, and theoretically prove that it is an unbiased approximation of the original method with an acceptable margin of error.
\item  The numerical experiments show our proposed model is superior to leading CF models and exhibits proficiency in modeling long-tail nodes.
\end{itemize}

%% file: sections/preliminaries.tex
\section{Preliminaries of Hyperbolic Geometry}
In this section, we briefly introduce concepts related to hyperbolic geometry, which builds up the foundation of our method.

\textbf{Minkowski (pseudo) Inner Product.} Consider the bilinear map
$
\langle \cdot, \cdot \rangle_{\mathcal{M}} : \mathbb{R}^{n+1} \times \mathbb{R}^{n+1} \rightarrow \mathbb{R}
$
defined by
\[
\langle \mathbf{u}, \mathbf{v} \rangle_{\mathcal{M}} := -u_0v_0 + \sum_{i=1}^{n} u_iv_i = \mathbf{u}^T \mathbf{J} \mathbf{v},
\]
where \( \mathbf{J} = \text{diag}(-1, 1, \ldots, 1) \in \mathbb{R}^{(n+1) \times (n+1)} \). It is called Minkowski (pseudo) inner product on \( \mathbb{R}^{n+1} \).
Given a constant \(K > 0\), the equality
$
\langle x, x \rangle_{\mathcal{M}} = -K
$
implies that
\[
x_0^2 = K + \sum_{i=1}^{n} x_i^2 \geq K.
\]
Then, \[
\|\mathbf{u}\|_{\mathcal{M}} = \sqrt{\langle \mathbf{u}, \mathbf{u} \rangle_{\mathcal{M}}}
\] 
is a well-defined norm. 

\textbf{Hyperbolic manifold \(\mathcal{H}^{n,K}\) and Tangent space \( T_x\mathcal{H}^{n,K} \).}
Consider the subset of $\mathbb{R}^{n+1}$ defined as follows:
\begin{align*}
\mathcal{H}^{n,K} &:= \{ \mathbf{x} \in \mathbb{R}^{n+1} : \langle \mathbf{x}, \mathbf{x} \rangle_{\mathcal{M}} = -K \text{ and } x_0 > 0 \}.
\end{align*}
Given any constant \( K > 0 \), the set \( \mathcal{H}^{n,K} \) is an $n$-dimensional embedded submanifold of \( \mathbb{R}^{n+1} \) with tangent space:
\begin{align*}
T_x\mathcal{H}^{n,K} &= \ker \mathrm{D}h(\mathbf{x})
= \{ \mathbf{u} \in \mathbb{R}^{n+1} : \langle \mathbf{x}, \mathbf{u} \rangle_{\mathcal{M}} = 0 \} \nonumber,
\end{align*}
which is an \( n \)-dimensional subspace of \( \mathbb{R}^{n+1} \).


\textbf{North Pole Point on \(\mathcal{H}^{n,K}\).} The point \( \mathbf{o} := (\sqrt{K}, 0, \ldots, 0) \in \mathcal{H}^{n,K} \) is called the north pole point of \( \mathcal{H}^{n,K} \).
We observe that
\begin{align*}
T_o\mathcal{H}^{n,K} &= \{ \mathbf{u} \in \mathbb{R}^{n+1} : \langle \mathbf{o}, \mathbf{u} \rangle_\mathcal{M} = -\sqrt{K}, u_0 = 0 \}
\cong \mathbb{R}^n.
\end{align*}

\textbf{Riemannian Distance on \(\mathcal{H}^{n,K}\).} The distance function induced by the Riemannian metric \(\langle \cdot, \cdot \rangle_{\mathcal{M}}\) is
\[
d^{K}_{\mathcal{M}}(\mathbf{x}, \mathbf{y}) = \sqrt{K} \operatorname{arcosh}\left(-\frac{\langle \mathbf{x}, \mathbf{y} \rangle_{\mathcal{M}}}{K}\right)
\]
for all \(x, y \in \mathcal{H}^{n,K}\).

\textbf{Exponential and Logarithmic Maps.} 
For $\mathbf{x} \in \mathcal{H}^n_K, \mathbf{v} \in T_x\mathcal{H}^n_K$, and $\mathbf{y} \in \mathcal{H}^n_K$ with $\mathbf{v} \neq 0$ and $\mathbf{y} \neq \mathbf{x}$, the exponential and logarithmic maps can be defined as:
\begin{align}
\operatorname{Exp}_x^K(\mathbf{v}) = \cosh\left(\frac{\|\mathbf{v}\|_{\mathcal{M}}}{\sqrt{K}}\right) \cdot x + \sqrt{K} \sinh\left(\frac{\|\mathbf{v}\|_{\mathcal{M}}}{\sqrt{K}}\right) \cdot \frac{\mathbf{v}}{\|\mathbf{v}\|_{\mathcal{M}}}
\label{exp}
\end{align}
and
\begin{align}
    \operatorname{Log}_x^K(\mathbf{y}) &= \frac{d^K_\mathcal{M}(\mathbf{x}, \mathbf{y})}{\|\text{Proj}_x(\mathbf{y})\|_\mathcal{M}} \cdot \text{Proj}_x(\mathbf{y}).
    \label{log}
\end{align}
Let \( x^\mathcal{E} \in \mathbb{R}^n \) denote input Euclidean features. Let \( o := (\sqrt{K}, 0, \ldots, 0) \) denote the north pole in \( \mathcal{H}^n_K \), which we use as a reference point to perform tangent space operations. We interpret \( (0, x^\mathcal{E}) \) as a point in \( T_o\mathcal{H}^n_K \) to map vectors from Euclidean space to Hyperbolic manifold, we implement
\begin{align*}
x^\mathcal{H} &:= \operatorname{Exp}_o^K ((0, x^\mathcal{E})).
\end{align*}
Notice the position of the zero elements in o and \( (0, x^\mathcal{E}) \) as vectors of \( \mathbb{R}^n \).

\textbf{Hyperbolic Weighted Mean.}
Given a set of hyperbolic points: $\mathcal{N} \subset \mathcal{H}^{d+1, K}$. For each point $\mathbf{x}_{i} \in \mathcal{N}$ and its corresponding weight $w_i\in\mathbf{w}$. We define the centroid of these points $\mathbf{c}\in \mathcal{H}^{d+1, K}$ as to minimizes the problem:
$$
\mathbf{Centroid}(\mathcal{N},\mathbf{w})=\arg\min _{\mathbf{c} \in \mathcal{H}^{d+1, K}} \sum_{{\mathbf{x}_{i}}\in \mathcal{N}} w_i\left(d_{\mathcal{M}}^K(\mathbf{c},\mathbf{x}_{i})\right)^2
$$
the closed-form solution is given in :
\begin{equation}
\mathbf{Centroid}(\mathcal{N},\mathbf{w})= \sqrt{K}\frac{\sum_{\mathbf{x}_{i} \in \mathcal{N}} w_i\mathbf{x}_{i}}{|\ \ ||\sum_{\mathbf{x}_{i}\in \mathcal{N}} w_i\mathbf{x}_{i}||_\mathcal{M}|}. 
\label{weight mean}
\end{equation}
When $\mathbf{w}$ is not given, the centroid of these points is defined as: 
\begin{equation}
\mathbf{Centroid}(\mathcal{N})= \sqrt{K}\frac{\sum_{\mathbf{x}_{i} \in \mathcal{N}}\mathbf{x}_{i}}{|\ \ ||\sum_{\mathbf{x}_{i}\in \mathcal{N}}\mathbf{x}_{i}||_\mathcal{M}|}. 
\label{mean}
\end{equation}

\textbf{Hyperbolic Matrix Multiplication}. For a matrix $\mathbf{W}$ and a vector $\mathbf{x}$ in hyperbolic space. We adopt the definition of \cite{hgcn}, define the hyperboloid matrix multiplication as
\begin{equation}
\label{hyper linear}
\mathbf{W} \otimes^{K} \mathbf{x} := \operatorname{Exp}_o^K \left( \mathbf{W} \operatorname{Log}_o^K (\mathbf{x}^\mathcal{H}) \right).
\end{equation}

\textbf{Parallel Transport}.
If two points \( \mathbf{x} \) and \( \mathbf{y} \) on the hyperboloid \( \mathcal{H}^{n,K} \) are connected by a geodesic, then the parallel transport of a tangent vector \( v \in T_x\mathcal{H}^{n,K}\) to the tangent space \( T_y\mathcal{H}^{n,K}\) is

\begin{equation}
P_{\mathbf{x} \rightarrow \mathbf{y}}(\mathbf{v}) = \mathbf{v} - \frac{\langle \log_\mathbf{x}(\mathbf{y}), \mathbf{v} \rangle_{\mathcal{M}}}{(d_{\mathcal{M}}^K(\mathbf{x},\mathbf{y}))^2} (\log_\mathbf{x}(\mathbf{y}) + \log_\mathbf{y}(\mathbf{x})).
\label{para}
\end{equation}

%% file: sections/Hgformer.tex
\begin{table}[t!]
	\centering
	\caption{{{Notations used in this paper}}}
	\vspace{-0.cm}
	\renewcommand\arraystretch{1.3}
		\begin{threeparttable} 
			\begin{tabular}{p{2.5cm}<{\centering}|p{5.4cm}<{\centering}}
				\hline\hline
				Notation  & Description \\ \hline
$\mathcal{V}_u$ \& $\mathcal{V}_i$  &The user and item set\\ 
$N$  &Number of users \\   
$M$  &Number of items\\
$N_i$ & The neighbors of node $i$\\
$\mathcal{E}$ &Interactions between users and items\\
$\Theta$ &Model parameters\\
$d$  &Dimension of latent embeddings.\\  
$\textbf{u}^{\mathcal{E}}$ \& $\textbf{i}^{\mathcal{E}}$ &User\&item embeddings in Euclidean space.\\
$\textbf{u}$ \& $\textbf{i}$ & User\&item embeddings in hyperbolic manifold.\\
$\textbf{u}_k$ \& $\textbf{i}_l$ & $k$-th and $l$-th vector of $\textbf{u}$ \& $\textbf{i}$ \\
$\textbf{u}^{(l)}$ \& $\textbf{i}^{(l)}$ & User\&item embeddings after $l$-th layer of LHGCN\\
$\textbf{q}_i^{(h)}$ \& $\textbf{k}_j^{(h)}$\& $\textbf{v}_j^{(h)}$ & $i$-th query, $j$-th key and value of $h$-head\\
$\mathbf{u}^{local}$ \& $\mathbf{i}^{local}$ & User\&item embeddings after LHGCN\\
$\mathbf{u}^{global}$ \& $\mathbf{i}^{global}$ & User\&item embeddings of hyperbolic cross attention.\\ 
$w_{i,j}^{(h)}$ & Similarity score of $i$-th user and $j$-item of $h$-th head in the self-attention\\
$\hat{\mathbf{u}}_i^{(h)}$ & $i$-th user embedding of $h$-th head after Euclidean weighted sum.\\
$\mathbf{u}_i'^{(h)}$ & $i$-th user embedding of $h$-th head after hyperbolic weighted sum.\\
$m$ & Number of random features.\\
                    
                    \hline
				\hline
			\end{tabular}
	\end{threeparttable} 
	\label{notation}
	\vspace{-0.cm}
\end{table}

\section{Hgformer: A Hyperbolic Graph Transformer for Recommendation}
In this section, we will elaborate on our proposed method. All the notations we use in this paper are detailed in Table~\ref{notation}. We formally define our tasks as follows: \textbf{Input:} The interaction graph of users and items $\mathcal{G}=(\mathcal{V}_u,\mathcal{V}_i,\mathcal{E}), \mathcal{E}\subseteq\mathcal{V}_u\times\mathcal{V}_i$. \textbf{Output:} A learned function $\mathcal{F}=(u,i|\mathcal{G},\Theta)$, where $u\in{V}_u,i\in{V}_i$ and $\Theta$ denote the model parameters.

\begin{figure*}[!tb]
    \centering
    \includegraphics[width=1.0\textwidth]{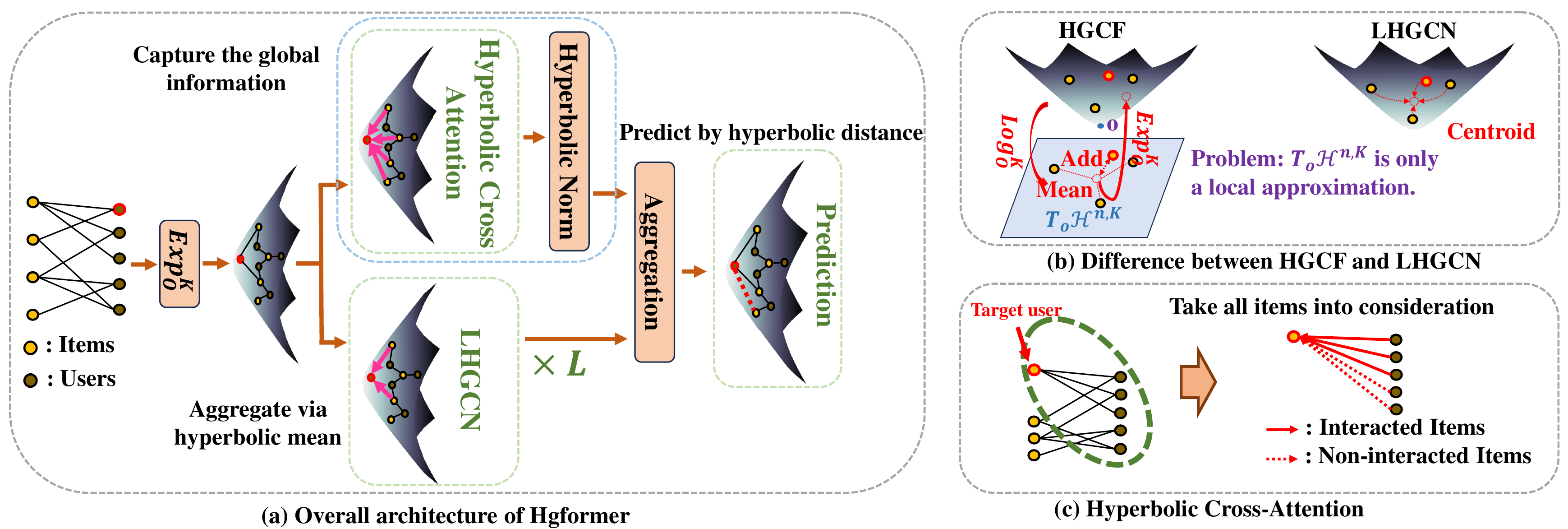}
    \caption{Figure(a) shows the overall architecture of Hgformer. We first map the user and item embeddings into a hyperbolic manifold. Then, we use several layers of LHGCN to capture the local information of the interaction graph and use a hyperbolic transformer to capture the overall information of the interaction graph. Finally, we combine these two information and make predictions on the hyperbolic manifold. Figure (b) shows the difference between HGCF and LHGCN in graph convolution. 
    In HGCF, embeddings are mapped from the hyperbolic manifold to the North Pole point’s tangent space for aggregation, then back to the hyperbolic manifold, causing information distortion due to the local approximation. LHGCN, on the other hand, aggregates information directly in the hyperbolic manifold.
    Figure (c) shows the process of hyperbolic cross-attention. Unlike GNN-based models that only consider items a user has interacted with, Hyperbolic Cross-Attention takes both the interacted and non-interacted items into account.}
    \label{fig:overall}
    \vspace{-5pt}
\end{figure*}

As is shown in Fig.~\ref{fig:overall}(a), in our framework, we first map the users and items into embedding space according to their IDs and use an exponential map to project the embeddings into hyperbolic space.
To capture the local structure of nodes in the user-item interaction graph, we design a Light Hyperbolic Graph Convolutional Network (LHGCN). On the other hand, to capture the global structure of the entire interaction graph, we propose a novel hyperbolic transformer, which is composed of a hyperbolic cross-attention mechanism with linear computation complexity and a hyperbolic normalization layer\cite{bdeir2023fully}.
In the final step, we aggregate the local structure information and the global information from both perspectives for prediction, optimized by a hyperbolic margin-ranking loss.

\subsection{Hyperbolic Embedding}
We first encode each user and item into the embedding space, which is denoted as ${\mathbf{u}^{\mathcal{E}}} = [{\mathbf{u}^{\mathcal{E}}_1};\ldots;{\mathbf{u}^{\mathcal{E}}_N]}$ and ${\mathbf{i}} = [{\mathbf{i}^{\mathcal{E}}_1};\ldots;{\mathbf{i}^{\mathcal{E}}_M}]$, where the superscript $\mathcal{E}$ means Euclidean space, $N$ and $M$ means the number of users and items. Then, we use an exponential map to project the embeddings into the hyperbolic space.
\begin{align*}\textstyle
\mathbf{u}^\mathcal{H}_k = \operatorname{Exp}_o^K ((0, \mathbf{u}^{\mathcal{E}}_k)), \quad \mathbf{i}^\mathcal{H}_k = \operatorname{Exp}_o^K ((0,\mathbf{i}^{\mathcal{E}}_k)).
\end{align*}
Since in the subsequent sections, we will only use the embeddings within the hyperbolic manifold, for convenience and to avoid confusion, we ignore the superscript $\mathcal{H}$ and simply denote the embeddings of users and items in the hyperbolic space as ${\mathbf{u}} = [{\mathbf{u}_1};\ldots;{\mathbf{u}_N]}$ and ${\mathbf{i}} = [{\mathbf{i}_1};\ldots;{\mathbf{i}_M}]$.

\subsection{LHGCN: Light Hyperbolic Graph Convolutional Networks}\label{LHGCN}
In CF tasks, each node (user or item) is represented by a unique ID, which lacks concrete semantics beyond being an identifier. In such scenarios, LightGCN~\cite{lightgcn} empirically demonstrated that performing multiple layers of nonlinear feature transformation does not provide benefits and increases the difficulty of model training. Therefore, removing nonlinear feature transformation from GCNs is a well-accepted approach in CF. 
To perform message-passing in hyperbolic space, a straightforward solution is to first project embeddings in the hyperbolic manifold back to the tangent space at the north pole point, performing parameter-free graph convolution, and then map them back to the hyperbolic manifold~\cite{hgcf,hrcf,hicf}.
However, since the tangent space at the north pole point is merely a local approximation of the north pole point~\cite{boumal2023intromanifolds} this can cause a certain degree of information loss. For this sake, we design a simple but efficient graph convolution method tailed for CF called LHGCN. Similar to HGCF~\cite{hgcf} and LightGCN~\cite{lightgcn}, LHGCN does not have trainable parameters and all computations are performed entirely on the hyperbolic manifold, eliminating the need for transformations between the hyperbolic manifold and Euclidean space. Specifically, we adopted hyperbolic centroid which is defined in Eq.~\ref{weight mean} to aggregate the messages of neighbors:
\begin{align*}\textstyle
{\mathbf{u}_i}^{(l+1)} &= \mathbf{Centroid}(\{{\mathbf{u}_i}^{(l)},\{{\mathbf{i}_k}^{(l)}:k\in N_{i}\}\}),\\
{\mathbf{i}_j}^{(l+1)} &= \mathbf{Centroid}(\{{\mathbf{i}_j}^{(l)},\{{\mathbf{u}_k}^{(l)}:k\in N_{j}\}\}),
\end{align*}
where $N_i$ denotes the neighbors of node $i$. Then the outputs of LHGCN are 
\begin{align*}\textstyle
\mathbf{u}^{local}=\mathbf{u}^{(L)} \text{ and } \mathbf{i}^{local}=\mathbf{i}^{(L)},
\end{align*}
where $L$ denotes the number of layers of LHGCN.

\subsection{Hyperbolic Transformer Model}
In this section, to address LHGCN's limitations in capturing the global information of the interaction graph and taking into account the unique structure of the bipartite graph in CF, we design a novel hyperbolic cross-attention mechanism for modeling global user-item interactions. Furthermore, since this cross-attention requires operating on all user-item pairs, with computational complexity $O(M\cdot N)$, we propose an approximation approach to reduce the computational complexity to $O(M+N)$.
\subsubsection{Hyperbolic Cross-Attention}
In CF tasks, there are only interactions between the user set and the item set, which form a bipartite graph. Intuitively, modeling the inner interactions among user groups or item groups would introduce noisy signals and thus deteriorate the performance \cite{min2022neighbour}. For this reason, we introduce a cross-attention mechanism to model only all possible user-item interactions, as detailed in the structure presented in Fig.~\ref{fig:overall}(c). Taking the $i$-th user vector as an example, firstly, the correlation between the $i$-th  user vector ($i \in \{1,\ldots, N\}$) and the $j$-th item vector ($j \in \{1,\ldots, M\}$) under a specific attention head $h$ is defined as:
\begin{align}
\label{basic_attenion}
    w^{(h)}_{i,j} &= \frac{\text{exp}\left(\operatorname{Sim}(\mathbf{q}_i^{(h)},\mathbf{k}_j^{(h)})/\tau\right)}{\sum_{l=1}^M \text{exp}\left(\operatorname{Sim}(\mathbf{q}_i^{(h)},\mathbf{k}_l^{(h)})/\tau\right)},
\end{align}
where $\mathbf{q}_i^{(h)}=\mathbf{W}_\text{Q}^{(h)}\otimes^{K}\mathbf{u}_i$, $\mathbf{k}_i^{(h)}=\mathbf{W}^{(h)}_\text{K}\otimes^{K}\mathbf{i}_i$ and $\otimes^{K}$ denotes hyperbolic matrix multiplication, which is defined in Eq.~\ref{hyper linear}. $\tau$ is the temperature parameter. $\operatorname{Sim}(\cdot,\cdot)$ is a function to calculate the similarity of two vectors in the hyperbolic manifold, and 
it was generally defined as:
\begin{align}
\label{similarity}
     \operatorname{Sim}(\mathbf{x},\mathbf{y})=f(-c_1d_{\mathcal{M}}^K(\mathbf{x},\mathbf{y})+c_2),
\end{align}
where $f(\cdot)$ is a monotonically increasing function,
such as exponential maps, linear functions, tanh, sigmoid, etc.. Then, the representation of the $i$-th user is updated by aggregating all item embeddings  with weights $\alpha_{i,j}$ :
\begin{align}
\label{euc agg}
      \hat{\mathbf{u}}_i^{(h)}= \sum_{j=1}^M w^{(h)}_{i,j}\mathbf{v}^{(h)}_j,
\end{align}
where $\mathbf{v}_j^{(h)}=W^{(h)}_\text{V}\otimes^{K}\mathbf{i}_j$. 

To ensure that the embedding stays in the hyperbolic manifold, we need an extra coefficient $c$ to scale the embedding according to Eq.~\ref{weight mean}, 
\begin{align}
      \mathbf{u}_i'^{(h)}= c\hat{\mathbf{u}}_i^{(h)},
\end{align}
where $c= \frac{K}{{|\ \ ||\hat{\mathbf{u}}_i^{(h)}||_\mathcal{M}|}}$ and $K$ is the curvature of the hyperbolic manifold. After that, we aggregate the embeddings of different heads by the hyperbolic centroid defined in Eq.~\ref{mean}:
\begin{equation}
\label{aggregation}
    \mathbf{u}^{global}_i= \mathbf{Centroid}(\mathbf{u}_i'^{(1)};\ldots;\mathbf{u}_i'^{(h)}).
\end{equation}
We calculate all item embeddings $\mathbf{i}^{global}_j, j\in\{1,\ldots,M\}$ in the same way.

For all the embeddings $\mathbf{x}=[\mathbf{u}^{global},\mathbf{i}^{global}]$.
Finally, for numerical stability, we adopted the definition of Hyperbolic Normalization from \cite{bdeir2023fully} and applied it to normalize the final embeddings $\mathbf{x}$:
\begin{equation}
\text{HN}(x) = \exp_{\beta}^{K} \left( \text{PT}_{\mathbf{o} \rightarrow \beta}^{K} \left( \gamma \cdot \frac{\text{PT}_{\mu \rightarrow 0}^{K} \left( \log_{\mu}^{K}(\mathbf{x}) \right)}{\sqrt{\sigma^2 + \epsilon}} \right) \right),
\end{equation}
where $\mu=\mathbf{Centroid}(\mathbf{x})$, which is Centroid of $\mathbf{x}$ in hyperbolic manifold and $\sigma^2=\frac{1}{M+N} \sum_{i=1}^{M+N} (d_{\mathcal{M}}^K(\mathbf{x}_i, \mu))^2$, which is the variance in hyperbolic space. And $\beta$ and $\epsilon$ are trainable vectors.

\subsubsection{Towards Linear Complexity}
\begin{figure}[!tb]
    \centering
    \includegraphics[width=0.5\textwidth]{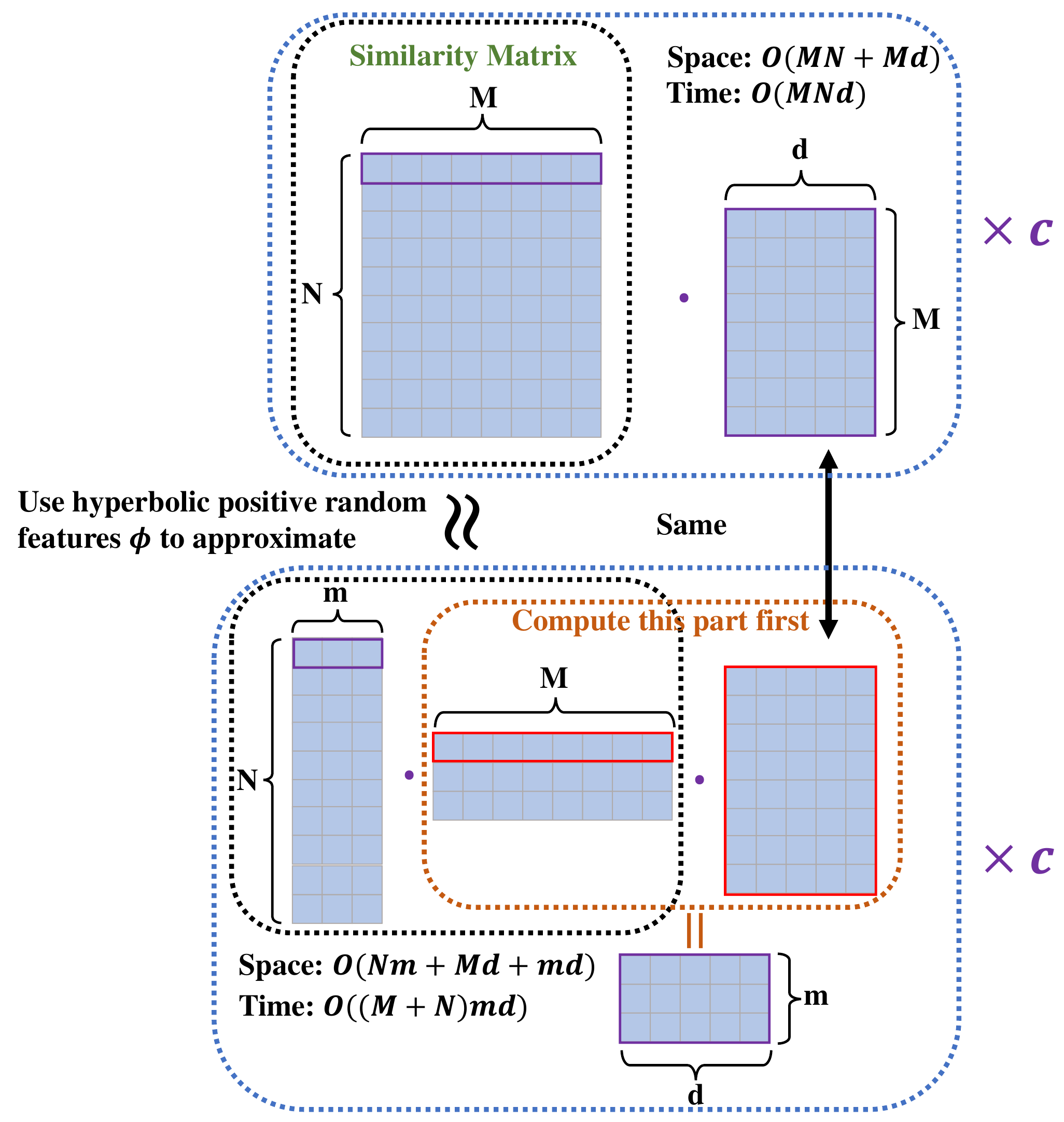}
    \caption{This figure shows how we reduce the complexity of hyperbolic graph transformers to linear. We first use $\phi$ to transform the similarity matrix into the multiplication of two smaller matrices and then, we change the order of computation order to reduce the computation complexity.}
    \vspace{-5mm}
    \label{fig:linear}
\end{figure}

In this section, we introduce the hyperbolic self-attention with only linear computational complexity. Although the above Hyperbolic cross-attention mechanism can model the global interactions between all users and items, its quadratic time complexity prevents its application in real-world scenarios when there are numerous users or items. To enable the application of Hyperbolic Transformers to larger datasets, as Fig. \ref{fig:linear} shown, the previous complexity of the similarity matrix of hyperbolic vectors is reduced to only $O((M+N)md)$ by our mechanism, where the dimensions $m$ and $d$ are much smaller than $M$ and $N$. 

Since the most computationally intensive part of the model is Eq.~\ref{euc agg}, in this section, our goal is to reduce the computational complexity of Eq.~\ref{euc agg} to linear.
Firstly, we redefined Eq.~\ref{similarity}.
Since $K>0$, $\operatorname{arcosh}(\cdot)$ is a monotonically increasing function, both $\left(-d_\mathcal{M}^K(\cdot, \cdot)\right)$ and the Minkowski inner product $\left\langle \cdot,\cdot \right\rangle_{\mathcal{M}}$ could be used to compare the similarity between different vectors in hyperbolic space. Then for $\mathbf{x},\mathbf{y}\in\mathcal{H}^{d+1}$, we redefine the similarity function $\operatorname{Sim}(\cdot,\cdot)$ as the Hyperbolic SoftMax similarity function, $\operatorname{HSM}(\cdot,\cdot):\mathcal{H}^{d+1,K} \times \mathcal{H}^{d+1,K} \rightarrow \mathbb{R}$:

$$\operatorname{HSM}(\mathbf{x}, \mathbf{y})\triangleq\exp\left(\left<\mathbf{x}, \mathbf{y}\right> _{\mathcal{M}}\right)$$
then Eq.~\ref{euc agg} is redefined as:
\begin{equation}
\label{Haggregation}
\hat{\mathbf{u}}_i^{(h)}=\sum_{n=1}^M\frac{\exp \left(\left<\mathbf{q}_i^{(h)},\mathbf{k}_j^{(h)}\right>_{\mathcal{M}}/\tau\right)}{\sum_{l=1}^M \exp \left(\left<\mathbf{q}_i^{(h)},\mathbf{k}_l^{(h)}\right>_{\mathcal{M}}/\tau\right)}
\cdot\mathbf{v}_{n}^{(h)},
\end{equation}
where $\mathbf{q}_i^{(h)}$, $\mathbf{k}_i^{(h)}$ and $\mathbf{v}_i^{(h)}$ follows the settings of Eq.~\ref{basic_attenion}.
Then, we use an estimation $\widetilde{\kappa}(\mathbf{q}_i^{(h)},\mathbf{k}_j^{(h)})$ to approximate $\exp \left(\left<\mathbf{q}_i^{(h)},\mathbf{k}_j^{(h)}\right>_{\mathcal{M}}\right)$ in Eq.~\ref{Haggregation} and we introduce Theorem 4.1, which proves that the aforementioned estimation is an unbiased estimation.

\noindent\textbf{Theorem 4.1}. \textit{For} $\mathbf{x}, \mathbf{y} \in \mathcal{H}^{d+1,K}$, \textit{with} 
$\mathbf{x} = (x_0, x_1, \cdots, x_d)^{\top}$, 
$\mathbf{y} = (y_0, y_1, \cdots, y_d)^{\top}$, 
\textit{and} 
$\widetilde{\mathbf{x}} = (x_1, x_2, \cdots, x_d)^{\top}$, 
$\widetilde{\mathbf{y}} = (y_1, y_2, \cdots, y_d)^{\top}$, 
\textit{we have an estimation function} 
$\widetilde{\kappa}(\cdot, \cdot) : \mathcal{H}^{d+1,K} \times \mathcal{H}^{d+1,K} \rightarrow \mathbb{R}$:
\begin{align}
\label{unbias}
\widetilde{\kappa}(\mathbf{x}, \mathbf{y}) = \exp & \left( \frac{-(x_0 + y_0)^2 + 2K}{2} \right) \notag \\ 
& \cdot\mathbb{E}_{\boldsymbol{\omega} \sim \mathcal{N}(\mathbf{0}_d, \mathbf{I}_d)} \left[ \exp (\boldsymbol{\omega}^{\top}(\widetilde{\mathbf{x}} + \widetilde{\mathbf{y}})) \right]
\end{align}
\textit{where $\boldsymbol{\omega} \sim \mathcal{N}\left(\boldsymbol{0}_d, I_d\right)$ ,}\textit{$\widetilde{\kappa}(\cdot, \cdot)$ is an unbiased estimation of the HSM function:}
\begin{equation}
    \widetilde{\kappa}(\mathbf{x}, \mathbf{y})=\operatorname{HSM}(\mathbf{x}, \mathbf{y})
\end{equation}
The proof of this theorem is given in Section~\ref{proof}.

Then, such an unbiased estimation function (Eq.~\ref{unbias}) can be converted into a dot product of vector functions approximately; the method of converting it is akin to kernel tricks shown as the following lemma: 

\noindent\textbf{Lemma 4.2}. \textit{Define the hyperbolic positive random features $\phi(\cdot):\mathcal{H}^{d+1,K}\rightarrow \mathbb{R}^m $ :}
$$
\phi(\mathbf{x})=\frac{\exp \left(\frac{K-x_0^2}{2}\right)}{\sqrt{m}}\left[\exp \left(\mathbf{\boldsymbol{\omega}}_1^{\top} \widetilde{\mathbf{x}}\right), \cdots, \exp \left(\mathbf{\boldsymbol{\omega}}_m^{\top} \widetilde{\mathbf{x}}\right)\right]
$$
\textit{where $\boldsymbol{\omega}_k \sim \mathcal{N}\left(\boldsymbol{0}_d, I_d\right)$ is i.i.d., $m$ is a constant that could be chosen smaller than $d$.} 
Then, we have:
\begin{equation}
\label{2}
    \phi(\mathbf{x})^{\top} \phi(\mathbf{y})\approx\widetilde{\kappa}(\mathbf{x}, \mathbf{y})=\operatorname{HSM}(\mathbf{x}, \mathbf{y}).
\end{equation}
The proof of this lemma is given in Section~\ref{proof}. 
We adopted a positive random feature map $\phi(\cdot):\mathcal{H}^{d+1,K}\rightarrow \mathbb{R}^m $ to approximate $\operatorname{HSM}$ function:
\begin{equation}
\label{1}
\begin{aligned}
     \operatorname{HSM}(\mathbf{x}, \mathbf{y})/\tau &\approx\phi(\frac{\mathbf{x}}{\sqrt{\tau}})^{\top} \phi(\frac{\mathbf{y}}{\sqrt{\tau}})) 
\end{aligned}
\end{equation}
 Eq.~\ref{1} is proved in Lemma 4.2 and for $\mathbf{x}\in \mathcal{H}^{d+1, K}$, $ \mathbf{x}=(x_0,x_1\cdots x_d)^{\top}$, $\widetilde{\mathbf{x}}=(x_1,x_2\cdots x_d)^{\top}$, the explicit form of positive random feature map with temperature parameter $\tau$ is defined as:
$$
\phi(\frac{\mathbf{x}}{\sqrt{\tau}})=\frac{\exp \left(\frac{K-x_0^2}{2\tau}\right)}{\sqrt{m}}\left[\exp \left(\frac{\mathbf{\boldsymbol{\omega}}_1^{\top} \widetilde{\mathbf{x}}}{\sqrt{\tau}}\right), \cdots, \exp \left(\frac{\mathbf{\boldsymbol{\omega}}_m^{\top} \widetilde{\mathbf{x}}}{\sqrt{\tau}}\right)\right].
$$
Then, we can change the computation order and extract common factors by using Eq.~\ref{2} to convert the HSM function into the dot product of two feature functions. Then the approximating aggregation function Eq.~\ref{approximation} only has linear complexity and the process is visualized in Fig.~\ref{fig:linear}. Subsequently, the final form of the aggregation function is proposed as follows:
\begin{equation}
\label{approximation}
    \begin{aligned}
        \hat{\mathbf{u}}_i^{'(h)}
        & \approx\sum_{j=1}^M \frac{\phi\left(\mathbf{q}_i/\sqrt{\tau}\right)^{\top}\phi\left(\mathbf{k}_j/\sqrt{\tau}\right)}{\sum_{n=1}^M \phi\left(\mathbf{q_i}/\sqrt{\tau}\right)^{\top}\phi\left(\mathbf{k}_n/\sqrt{\tau}\right)} \cdot \mathbf{v}_j\\
        &=\frac{\phi\left(\mathbf{q}_i/\sqrt{\tau}\right)^{\top} \sum_{j=1}^M \phi\left(\mathbf{k}_j/\sqrt{\tau}\right) \cdot\left(\mathbf{v}_j\right)^{\top}}{\phi\left(\mathbf{q}_i/\sqrt{\tau}\right)^{\top} \sum_{k=1}^M \phi\left(\mathbf{k}_n/\sqrt{\tau}\right)}.
    \end{aligned}
\end{equation}
The error of approximation is bounded and we have: 

\noindent\textbf{Theorem 4.2}: \textit{The error function of approximation $$\Delta=\left|\operatorname{HSM}(\mathbf{x}, \mathbf{y})-\phi(\mathbf{x})^{\top} \phi(\mathbf{y})\right|$$
is bounded by $\mathcal{O}\left(\sqrt{\frac{\exp (3( \delta-K) )}{m \epsilon}}\right)$ with the probability that:}
\begin{equation}
   \mathbb{P}(\Delta \leqslant \sqrt{\frac{\exp (3 (\delta-K))}{m \epsilon}}) \leq 1 - \epsilon
\end{equation}
\textit{assuming that $\|\mathbf{x}\|^2_{\mathcal{E}}\leq\delta$ ,$\|\mathbf{y}\|^2_{\mathcal{E}}\leq\delta$ for $\mathbf{x}, \mathbf{y}\in\mathcal{H}^{d+1,K}$.}
The proof is given in Section~\ref{proof}.

Since the upper bound of the error function depends only on the Euclidean norm $\delta$, the curvature constant $K$, the number of positive random features $m$, and the demanding error accuracy $\epsilon$, we can reduce the error by normalizing the vectors and process them in a suitable hyperbolic space, or increase $m$.

\subsection{Embedding Aggregation and Optimization}

\noindent\textbf{Embedding Aggregation.}
To aggregate both structural and global information, we map the embeddings back to Euclidean space using the Log function, and then perform a weighted average:
\begin{equation}
\begin{aligned}
\label{aggregation1}\mathbf{u}^{final}&=\operatorname{Exp}^K_o(\alpha\operatorname{Log} ^K_o(\mathbf{u}^{global})+(1-\alpha)\operatorname{Log} ^K_o(\mathbf{u}^{local})) \\
\mathbf{i}^{final}&=\operatorname{Exp}^K_o(\alpha\operatorname{Log} ^K_o(\mathbf{i}^{global})+(1-\alpha)\operatorname{Log} ^K_o(\mathbf{i}^{local}))
\end{aligned}
\end{equation}
where $\alpha$ is a hyperparameter between 0 and 1. 

\noindent\textbf{Prediction.}
Margin ranking loss has been extensively used in recommendation tasks \cite{hgcf}, which separates positive and negative pairs of user items by a given margin. When the gap between a negative and a positive user-item pair exceeds this margin, neither pair contributes to the overall loss, enabling the optimization process to focus on the difficult pairs in the data set. In this work, we use the hyperbolic version of margin-ranking loss as prediction loss. The prediction loss is defined as:
\vspace{-2mm}

\begin{equation*}
\begin{split}
L(\mathbf{u}^{final},\mathbf{i}_{neg}^{final},\mathbf{i}_{pos}^{final}) = \max \bigg( d_{\mathcal{M}}(\mathbf{u}^{final}, \mathbf{i}_{pos}^{final})^2 
\\- d_{\mathcal{M}}(\mathbf{u}^{final}, \mathbf{i}_{neg}^{final})^2 + \lambda , 0 \bigg)
\end{split}
\end{equation*}
where $\lambda$ is a non-negative hyperparameter. $\mathbf{i}_{pos}^{final}$ are the embeddings of the positive samples of this user and $\mathbf{i}_{neg}^{final}$ are the embeddings of negative samples of this user in the same hyperbolic manifold. Positive samples refer to the items that the user has interacted with, while negative samples refer to randomly sampled items that the user has not interacted with.

%% file: sections/experiments.tex
\begin{table*}[ht]
\centering
\caption{Overview of performance. N@10 and R@10 are abbreviations for the metrics NDCG@10 and Recall@10, respectively. * represents the significance level $p$-value $<0.05$. The highest scores for each dataset and metric are emphasized in bold, while the second-best ones are underlined.}
\label{results}
\resizebox{\textwidth}{!}{%
\begin{tabular}{@{}>{\raggedright\arraybackslash}p{2.5cm}lccccccccccc@{}}
\toprule
\textbf{Dataset}         & \textbf{Metric} & \textbf{BPR} & \textbf{NGCF} & \textbf{LightGCN} & \textbf{HMLET} & \textbf{HGCF} & \textbf{HICF} & \textbf{HRCF} & \textbf{SGFormer} & \textbf{NodeFormer} & \textbf{Hypformer} & \textbf{Hgformer} \\ \midrule
\multirow{4}{*}{\textbf{Amazon Book}} &\textbf{Recall@10}     & 0.0357 & 0.0427 & 0.0581 & 0.0347 & 0.0748 & \underline{0.0766} & 0.0745 & 0.0545 & 0.0379 & 0.0376 & \textbf{0.0901*} \\
                            & \textbf{Recall@20}      & 0.0538 & 0.0676 & 0.0910 & 0.0553 & 0.1089 & \underline{0.1086} & 0.1080 & 0.0841 & 0.0602 & 0.0605 & \textbf{0.1291*} \\
                            & \textbf{NDCG@10}        & 0.0218 & 0.0250 & 0.0344 & 0.0201 & 0.0467 & \underline{0.0485} & 0.0466 & 0.0332 & 0.0222 & 0.0215 & \textbf{0.0573*} \\
                            & \textbf{NDCG@20}        & 0.0267 & 0.0317 & 0.0433 & 0.0256 & 0.0560 & \underline{0.0571} & 0.0560 & 0.0412 & 0.0282 & 0.0277 & \textbf{0.0681*} \\ \midrule
\multirow{4}{*}{\textbf{Amazon CD}}   & \textbf{Recall@10}    & 0.0473 & 0.0755 & 0.0770 & 0.0601 & 0.0797 & \underline{0.0826} & 0.0824 & 0.0435 & 0.0454 & 0.0613 & \textbf{0.0977*} \\
                            & \textbf{Recall@20}      & 0.0726 & 0.1151 & 0.1174 & 0.0874 & 0.1184 & \underline{0.1211} & 0.1176 & 0.0695 & 0.0710 & 0.0936 & \textbf{0.1401*} \\
                            & \textbf{NDCG@10}        & 0.0268 & 0.0426 & 0.0437 & 0.0351 & 0.0451 & \underline{0.0489} & 0.0477 & 0.0242 & 0.0253 & 0.0340 & \textbf{0.0567*} \\
                            & \textbf{NDCG@20}        & 0.0334 & 0.0529 & 0.0542 & 0.0421 & 0.0552 & \underline{0.0580} & 0.0577 & 0.0310 & 0.0319 & 0.0424 & \textbf{0.0678*} \\ \midrule
\multirow{4}{*}{\textbf{Amazon Movie}} & \textbf{Recall@10}      & 0.0585 & 0.0580 & 0.0702 & 0.0491 & 0.0740 & 0.0740 & \underline{0.0761} & 0.0560 & 0.0338 & 0.0294 & \textbf{0.0803*} \\
                            & \textbf{Recall@20}      & 0.0929 & 0.0913 & 0.1106 & 0.0817 & 0.1110 & 0.1135 & \underline{0.1157} & 0.0871 & 0.0575 & 0.0528 & \textbf{0.1203*} \\
                            & \textbf{NDCG@10}        & 0.0362 & 0.0351 & 0.0440 & 0.0293 & 0.0464 & 0.0465 & \underline{0.0481} & 0.0343 & 0.0200 & 0.0150 & \textbf{0.0503*} \\
                            & \textbf{NDCG@20}        & 0.0455 & 0.0442 & 0.0549 & 0.0382 & 0.0566 & 0.0572 & \underline{0.0589} & 0.0428 & 0.0265 & 0.0210 & \textbf{0.0612*} \\ \midrule
\multirow{4}{*}{\textbf{Douban Book}} & \textbf{Recall@10}      & 0.1059 & 0.1280 & 0.1313 & 0.0926 & 0.1375 & \underline{0.1388} & 0.1357 & 0.0886 & 0.0777 & 0.0446 & \textbf{0.1462*} \\
                            & \textbf{Recall@20}      & 0.1588 & 0.1832 & 0.1906 & 0.1440 & 0.1935 & \underline{0.1938} & 0.1892 & 0.1349 & 0.1225 & 0.0772 & \textbf{0.2052*} \\
                            & \textbf{NDCG@10}        & 0.0706 & 0.0864 & 0.0922 & 0.0630 & 0.0960 & \underline{0.0968} & 0.0943 & 0.0610 & 0.0484 & 0.0289 & \textbf{0.1030*} \\
                            & \textbf{NDCG@20}        & 0.0850 & 0.1013 & 0.1078 & 0.0768 & 0.1113 & \underline{0.1117} & 0.1091 & 0.0734 & 0.0610 & 0.0380 & \textbf{0.1189*} \\ \midrule
\multirow{4}{*}{\textbf{Douban Movie}} & \textbf{Recall@10}     & 0.0616 & 0.1373 & 0.1339 & 0.1178 & 0.1348 & 0.1370 & \underline{0.1393} & 0.1259 & 0.1118 & 0.0678 & \textbf{0.1405*} \\
                            & \textbf{Recall@20}      & 0.0970 & 0.2042 & 0.1989 & 0.1802 & 0.1992 & 0.2034 & \underline{0.2036} & 0.1877 & 0.1722 & 0.1106 & \textbf{0.2068*} \\
                            & \textbf{NDCG@10}        & 0.0676 & 0.1256 & 0.1326 & 0.1227 & 0.1248 & 0.1297 & \underline{0.1342} & 0.1263 & 0.1002 & 0.0780 & \textbf{0.1322*} \\
                            & \textbf{NDCG@20}        & 0.0725 & 0.1389 & 0.1435 & 0.1325 & 0.1377 & 0.1424 & \underline{0.1456} & 0.1364 & 0.1133 & 0.0845 & \textbf{0.1447*} \\ \midrule
\multirow{4}{*}{\textbf{Douban Music}} & \textbf{Recall@10}     & 0.1084 & 0.1229 & 0.1258 & 0.0916 & 0.1218 & \underline{0.1276} & 0.1271 & 0.1071 & 0.0791 & 0.0313 & \textbf{0.1386*} \\
                            & \textbf{Recall@20}      & 0.1606 & 0.1803 & 0.1802 & 0.1418 & 0.1791 & \underline{0.1843} & 0.1791 & 0.1599 & 0.1239 & 0.0531 & \textbf{0.1955*} \\
                            & \textbf{NDCG@10}        & 0.0783 & 0.0899 & 0.0966 & 0.0727 & 0.0927 & \underline{0.0942} & 0.0940 & 0.0842 & 0.0534 & 0.0253 & \textbf{0.1024*} \\
                            & \textbf{NDCG@20}        & 0.0917 & 0.1045 & 0.1095 & 0.0840 & 0.1068 & \underline{0.1085} & 0.1070 & 0.0966 & 0.0656 & 0.0306 & \textbf{0.1165*} \\ \bottomrule
\end{tabular}%
}
\end{table*}

\section{Experiments}\label{evaluation}
In this section, we conduct extensive experiments on multiple public datasets to evaluate the proposed method and primarily address the following questions.
\begin{itemize}[\setlength{\itemindent}{\dimexpr\labelwidth+\labelsep}]
    \item[\textbf{RQ1:}] How does Hgformer perform compared to baselines?
    \item[\textbf{RQ2:}] How does each module contribute to the performance?
    \item[\textbf{RQ3:}] How well does Hgformer perform on the head and tail items?
     \item[\textbf{RQ4:}] Is Hgformer sensitive to different hyperparameters?
     \item[\textbf{RQ5:}] Why does Hgformer perform better than other models?
\end{itemize}

\subsection{Experimental Setup}
In this section, we will describe the settings of our experiments.

\subsubsection{Baselines}
To demonstrate the effectiveness of our proposed model, we compare our model with four categories of models: (A) Traditional models BPR\cite{rendle2012bpr} and DMF\cite{xue2017deep}. (B) GNN-based models LightGCN\cite{lightgcn} and HMLET. (C) Hyperbolic GNN-based methods HGCF\cite{hgcf}, HRCF\cite{hrcf} and HICF\cite{hicf}. (D) Graph Transformer-based methods SGFormer\cite{sgformer}, NodeFormer\cite{nodeformer}.

We introduce the baseline models in detail as follows:
\begin{itemize}
\item[$\bullet$] \textbf{BPR} \cite{rendle2012bpr}is a basic matrix factorization model that is trained in a pairwise way.
\item[$\bullet$] \textbf{NGCF} \cite{rendle2012bpr}is a GCN-based model that leverages high-order connectivity in the user-item graph by propagating embeddings and explicitly incorporating collaborative signals into the learned representations.
\item[$\bullet$] \textbf{LightGCN} \cite{lightgcn} is a state-of-the-art GCN model that eliminates feature transformations and non-linear activations deemed unnecessary for collaborative filtering in GCN.
\item[$\bullet$] \textbf{HMLET} \cite{kong2022linear} combines linear and nonlinear propagation layers for general recommendation and yields better performance.
\item[$\bullet$] \textbf{HGCF} \cite{hgcf} is a hyperbolic graph neural network model for recommendation. 

\item[$\bullet$] \textbf{HRCF} \cite{hrcf} is a hyperbolic graph neural network model with geometric regularization for recommendation.
\item[$\bullet$] \textbf{HICF} \cite{hicf} is a hyperbolic graph neural network model with geometric aware margin ranking learning for recommendation.
\item[$\bullet$] \textbf{SGFormer} \cite{sgformer} a linear graph transformer model for scalable graphs.
\item[$\bullet$] \textbf{NodeFormer} \cite{nodeformer} a kernel function-based graph transformer model for scalable graphs.
\item[$\bullet$] \textbf{Hypformer} \cite{nodeformer} a hyperboilic graph transformer model.

\end{itemize}

\subsubsection{Datasets}
To validate the effectiveness of the model, we conducted experiments on six datasets with varying sizes and densities, including Amazon datasets\footnote{\url{http://snap.stanford.edu/data/amazon}} (Amazon Book, Amazon CD, and Amazon Movie), as well as Douban datasets\footnote{\url{https://www.douban.com}} (Douban Book, Douban Movie, and Douban Music). Detailed statistics of the datasets are summarized in Table~\ref{statistics}.
\begin{table}[h]
    \caption{Dataset Statistics}
    \label{statistics}
    \centering
    \begin{tabular}{lcccc}
        \toprule
        Dataset & \#User & \#Item & \#Interactions & Density \\
        \midrule
        Amazon-Book &211,170  &163,789  &5,069,747  & 0.015\% \\
        Amazon-CD & 66,317 & 58,869 & 952,547 & 0.024\% \\
        Amazon-Movie & 26,969 & 18,564 & 762,957 & 0.150\% \\
        Douban-Book & 18,086 &33,068 &809,248 &0.140\% \\
        Douban-Movie &22,041 &25,802 &2,553,305  & 0.459\% \\
        Douban-Music &15,996  &39,749  &1,116,984  &0.176\% \\        
        \bottomrule
    \end{tabular}
\end{table}

\subsubsection{Evaluation Metrics}
We used two metrics for the evaluation. The first metric is Recall@K, which measures the fraction of relevant items retrieved out of all relevant items.
The second metric is NDCG@K, which is a measure of ranking quality in which positions are discounted logarithmically. It accounts for the position of the hits by assigning higher scores to hits at the top ranks. The formal definition is given as follows:
We employed two metrics for evaluation. The first metric is Recall@K, which measures the fraction of relevant items retrieved out of all relevant items, which is formally defined as:

\begin{equation}
\text{Recall@K} = \frac{1}{|\mathcal{U}|} \sum_{u \in \mathcal{U}} \frac{|\hat{R}(u) \cap R(u)|}{|R(u)|},
\end{equation}
where $\mathcal{U}$ is the set of all users, $\hat{R}(u)$ represent a ranked list of items that a model produces, and $R(u)$ represent a ground-truth set of items that user has interacted with.
The second metric is NDCG@K, which is a measure of ranking quality where positions are discounted logarithmically. It accounts for the position of the hits by assigning higher scores to hits at top ranks and it is formally defined as:
\begin{equation}
\begin{split}
\text{NDCG@K} = \frac{1}{|\mathcal{U}|} \sum_{u \in \mathcal{U}} \left( \frac{1}{\sum_{i=1}^{\min(|R(u)|, K)} \frac{1}{\log_2(i+ 1)}} \right. \\
\left. \sum_{i=1}^{K} \frac{\delta(i \in R(u))}{\log_2(i + 1)} \right),
\end{split}
\end{equation}
where \( \delta(\cdot) \) is an indicator function. In this research, we set K as 10 and 20.


\subsection{Overall Performance Comparison (RQ1)}
We compared Hgformer with traditional models, GNN-based models, hyperbolic GNN-based models, and graph transformer-based models across five datasets of different sizes, as shown in Table~\ref{results}. We find the following observations:

\begin{enumerate}
\item Across all datasets, Hgformer demonstrates consistent performance improvements. The improvements are particularly significant in the Amazon Book and Amazon CD datasets. On the Amazon Book dataset, compared to the second-best model, HICF, Hgformer achieves relative improvements of 17.6\%, 18.9\%, 18.1\%, and 19.3\% in Recall@10, Recall@20, NDCG@10, and NDCG@20, respectively. Similarly, on the Amazon CD dataset, the relative improvements are 18.3\%, 15.7\%, 15.9\%, and 16.9\%, respectively.

\item It can be seen that all three hyperbolic GCN-based models show noticeable improvements over traditional GNN models, demonstrating the advantages of hyperbolic GCN in handling recommendation tasks. This is attributed to the fact that the capacity of hyperbolic space exponentially increases with radius, aligning well with the power-law distributed user-item network.

\item We also compared our model with three graph transformer models (SGFormer, NodeFormer, and Hypformer). However, since these three models are designed for node classification tasks and compute attention for all node pairs, they tend to overemphasize the relationships between users and between items. This leads to poor performance in recommendation tasks.

\end{enumerate}

\begin{figure*}[!t]
    \centering
    \includegraphics[width=1.0\textwidth]{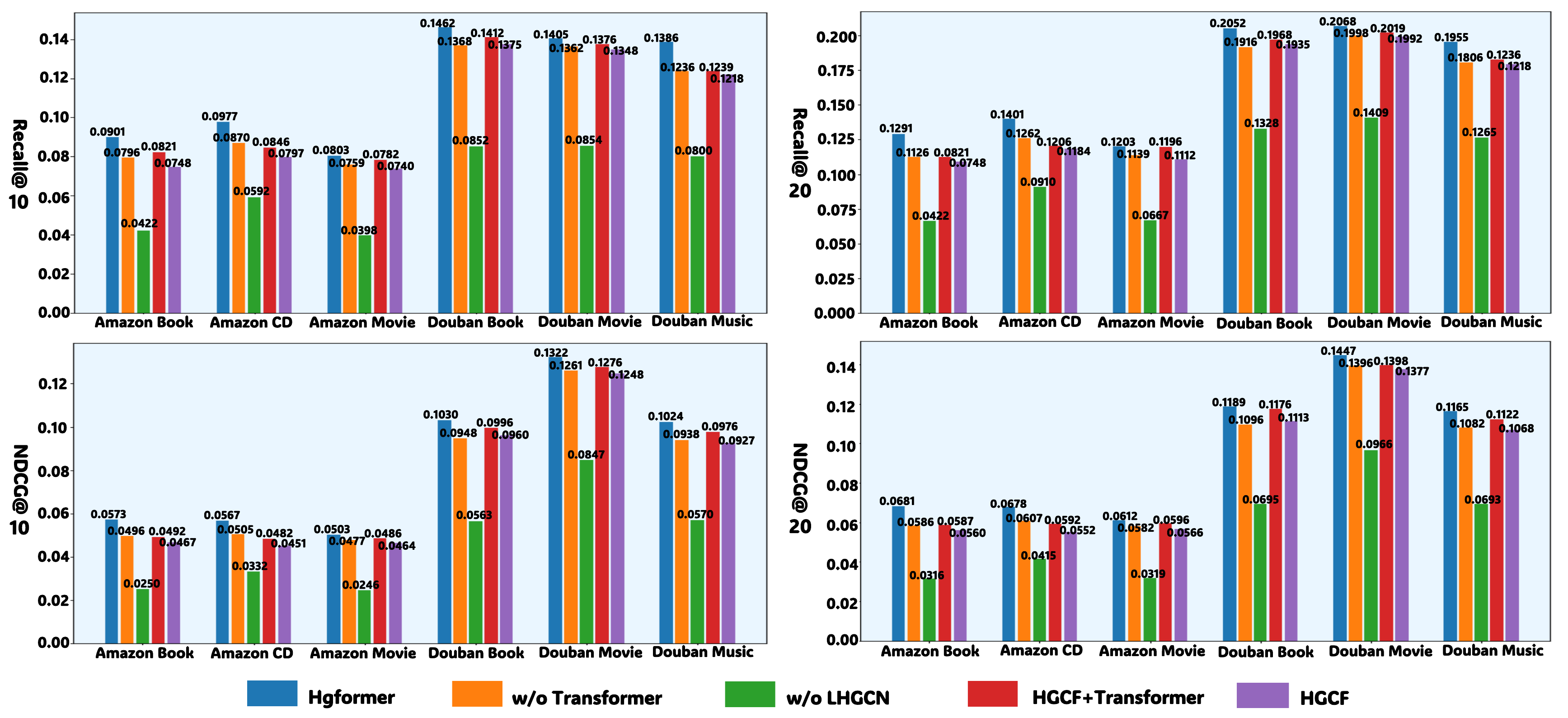}
    \caption{To validate the contribution of each module in the model, we individually removed the LHGCN and the Hyperbolic Transformer for evaluation. In the figure, 'w/o LHGCN' and 'w/o Transformer' represent the effects after removing the LHGCN and the Hyperbolic Transformer, respectively. Furthermore, to validate the effectiveness of LHGCN, we also compared LHGCN with HGCF. Specifically, we replaced LHGCN in Hgformer with HGCF and tested the performance and in the figure, it is noted as 'HGCF+Transformer'.}
    \label{fig.ablation}
\end{figure*}

\subsection{Ablation Analysis (RQ2)}
We conduct ablation studies on two main components of Hgformer. The results are shown in Fig.~\ref{fig.ablation}. We can observe that removing either the LHGCN or the Transformer component leads to a decline in the model's performance. Replacing LHGCN with HGCF also leads to a decline. To be specific, we have the following observations:

\begin{enumerate}
\item The removal of LHGCN results in the most significant drop in performance. This indicates that LHGCN plays a dominant role in the CF task, and solely using the Hyperbolic Transformer to capture global information between users and items, while ignoring the inherent topological structure of the existing interaction graph, is not sufficient to effectively capture potential user-item relations. Therefore, in recommendation tasks or link prediction tasks, it is difficult to achieve good results by completely abandoning GNNs and relying solely on transformers for prediction. A better strategy for link prediction and recommendation tasks is to use transformers as a supplementary tool.

\item Removing the Hyperbolic Graph Transformer also leads to a remarkable decline in model performance. Furthermore, we observed that incorporating the Hyperbolic Transformer led to certain improvements in HGCF, which demonstrates its effectiveness.

\item Replacing LHGCN with HGCF also results in a decline. Through the direct comparison between LHGCN and HGCF, we found that LHGCN outperforms HGCF on the majority of datasets (Amazon Book, Amazon CD, Amazon Movie, Douban Movie). The only exception is the Douban Book dataset, where LHGCN performs slightly worse than HGCF (Recall@10: 0.1368 vs. 0.1375; Recall@20: 0.1916 vs. 0.1935; NDCG@10: 0.0948 vs. 0.0960; NDCG@20: 0.1096 vs. 0.1113). As mentioned in Section~\ref{LHGCN}, HGCF requires mapping embeddings back to the tangent space at the North Pole point during message aggregation, which results in information distortion during this process. In contrast, LHGCN performs graph convolution entirely on the hyperbolic manifold and achieves better performance.

\end{enumerate}

\begin{figure*}[!t]
    \centering
\includegraphics[width=1\textwidth]{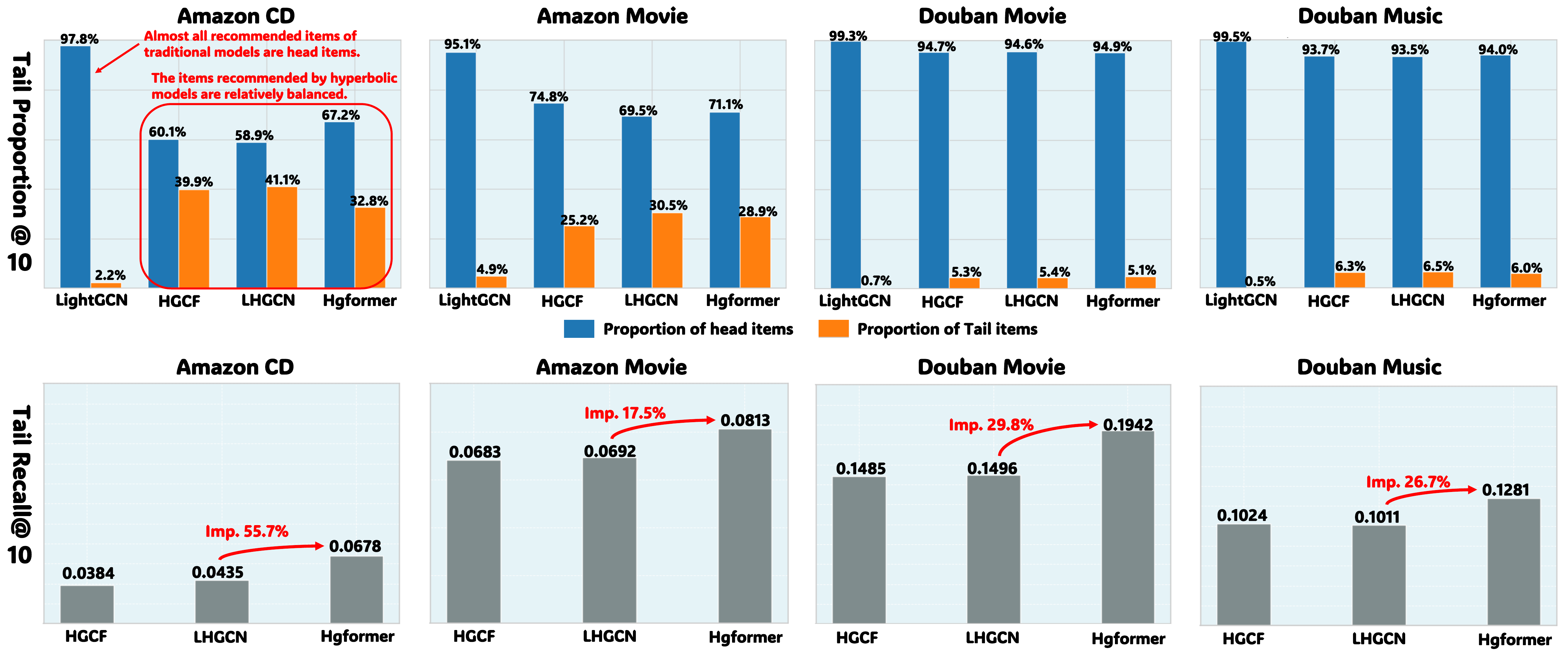}
    \caption{We define the top 20\% of items in terms of popularity as head items and the rest as tail items. We calculated the proportion of head items and tail items among all recommended items for each model.  To further analyze the performance of Hgformer on tail items, we evaluate the performance of three hyperbolic-based models on tail items.}
    \label{head_tail1}
    \vspace{-6mm}
\end{figure*}

\vspace{-3mm}
\subsection{Analysis on Tail Items (RQ3)}
In this section, we analyze the results on tail items to demonstrate Hgformer's capability to mitigate long-tail issues.

\subsubsection{Tail percentage analysis} 
We calculate the proportion of tail items recommended by each model and tail items are defined as those whose popularity ranks in the last 80\%. For this purpose, we designed a metric called tail percentage, which is formally defined as follows:
\begin{equation}
\text{TailPercentage@K} = \frac{1}{|U|} \sum_{u \in U} \frac{\sum_{i \in R_u} \delta(i \in T)}{|R_u|},
\vspace{-3mm}
\end{equation}
where $R_u$ is the set of items recommended to user $u$, $T$ is the set of tail items and $\sigma$ is an indicator function. This metric gives the proportion of head items and tail items among all the items recommended by the model. We conducted evaluations on the Amazon CD and Amazon Movie datasets. As shown in the upper part of Fig.~\ref{head_tail1}, it can be observed that the Euclidean space-based model (LightGCN) recommends almost only head items to users in both datasets (97.8\% in the Amazon CD dataset \& 95.1\% in the Amazon Movie dataset, 99.3\% in Douban Movie dataset and 99.5\% in Douban Music dataset). This indicates Euclidean space-based models tend to overlook tail items in CF tasks and aggravate the Matthew Effect. Conversely,  this phenomenon is significantly mitigated by the three hyperbolic space-based models, showing that hyperbolic space is more suitable for the long-tail setting in CF. It can be observed that the model's emphasis on tail items differs between the Amazon and Douban datasets, which is primarily due to their significant differences in the density and data distribution.

\subsubsection{Analysis of model's performance on tail items} 
To further investigate the performance of each hyperbolic-based model on tail items, in the second experiment, we evaluate the models' performance solely on tail items by calculating the recall@10 metric. As shown in Fig.~\ref{head_tail1}, Hgformer significantly outperforms the other two hyperbolic space-based models on tail items. This is because Hgformer not only leverages the hyperbolic manifold to capture the hierarchical structure of the data but also introduces hyperbolic cross-attention which is beneficial for capturing global information. This allows the model to gather more information for tail nodes during the message-passing process, thereby improving the accuracy of recommendation. 

\begin{figure*}[t!]
    \centering
    \includegraphics[width=1.0\textwidth]{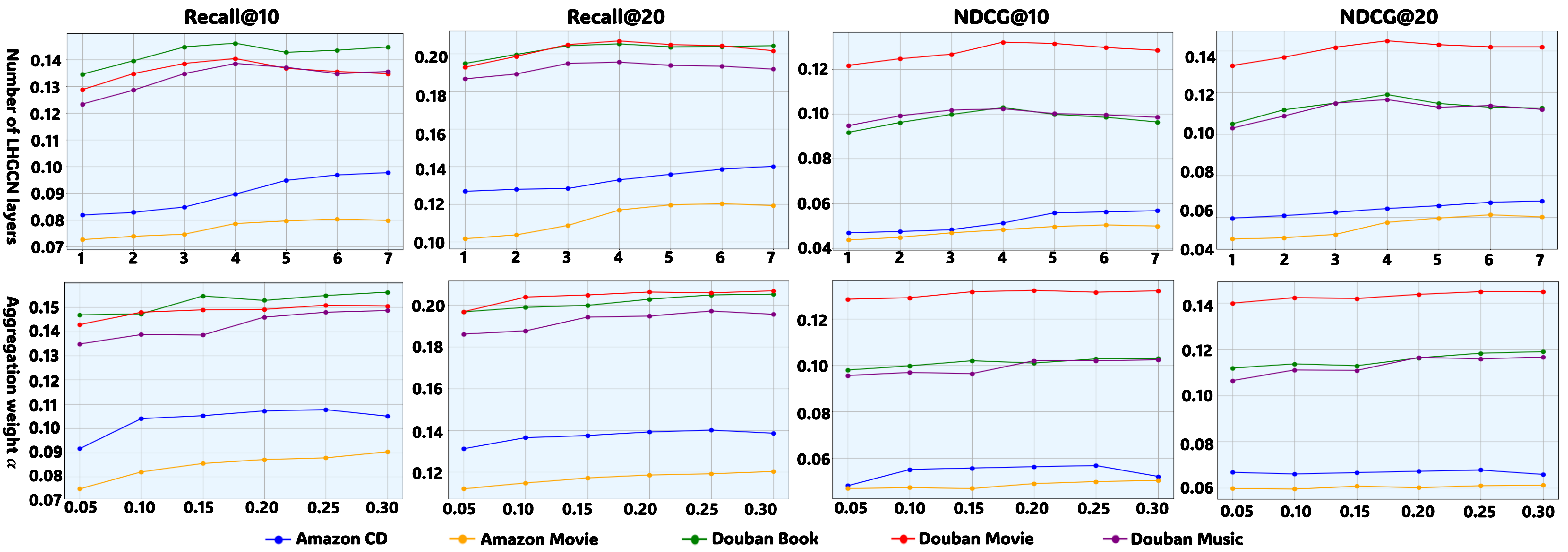}
    \caption{Sensitivity analysis on a number of LHGCN layers and aggregation weight $\alpha$}
    \label{sen_analysis}
\end{figure*}

\subsection{Sensitivity Analysis (RQ4)}
\label{sensitivity}
To evaluate the stability of our model, we conducted a sensitivity analysis on four hyperparameters of our model: number of LHGCN layers and aggregation weight $\alpha$. The results are shown in Fig.~\ref{sen_analysis}. 

\begin{enumerate}
\item Our model remains relatively stable within a certain range for both aggregation weight and the number of LHGCN layers.

\item In the sensitivity analysis of aggregation weights, our model demonstrated relative stability in the range of 0.2 to 0.3, achieving better performance within this interval.

\item For the number of LHGCN layers, we observe that the optimal performance range of the model differs between the Amazon dataset and the Douban dataset. For the Amazon dataset, the optimal number of LHGCN layers falls within the range of 6 to 7, whereas for the Douban dataset, the optimal performance is achieved roughly between 4 and 5 layers.
\end{enumerate}

\subsection{Case study (RQ5)}
\begin{figure}[!t]
    \centering
\includegraphics[width=0.5\textwidth]{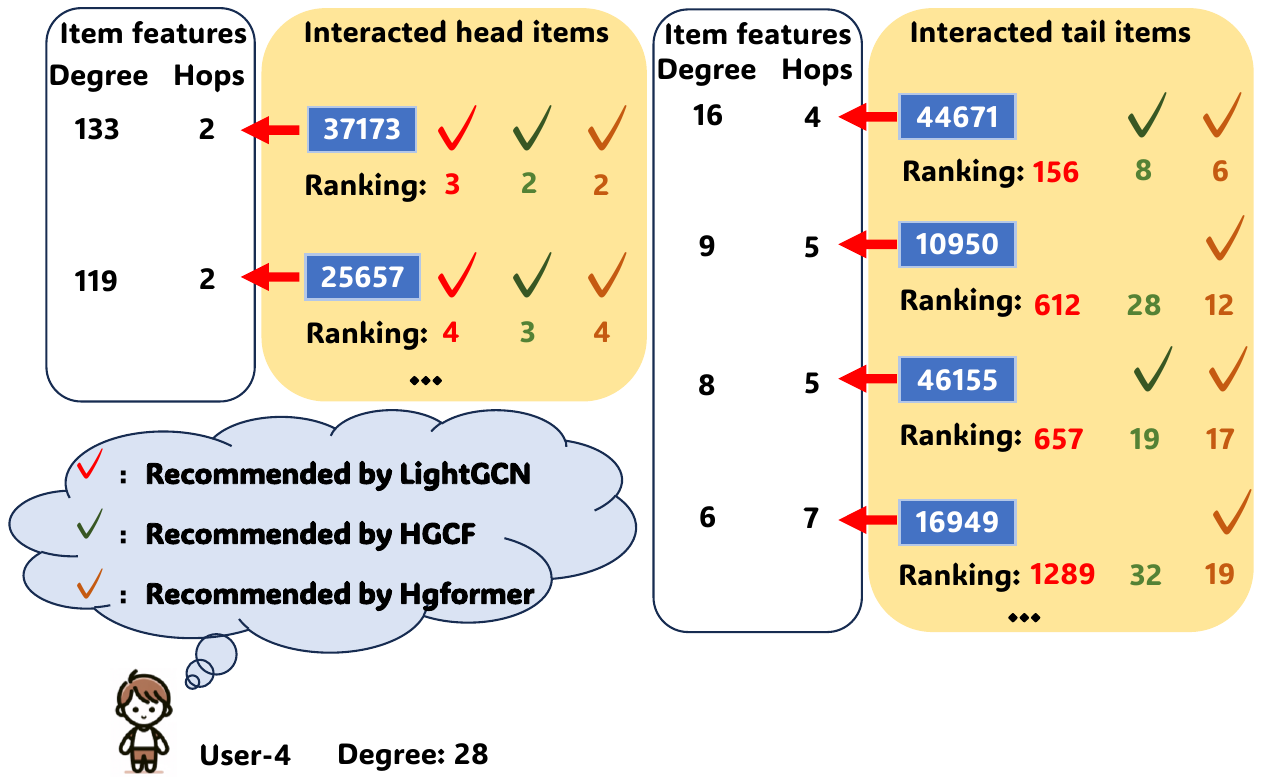}
    \caption{We selected user number 4 from the Amazon CD dataset as the subject of analysis. We analyzed the items recommended by Hgformer in the @20 task.}
    \label{head_tail}
    \vspace{-6mm}
\end{figure}

In this section, we present a case study to demonstrate the effectiveness of our model in addressing both the local structure modeling problem of GNN-based models and high distortion problem of models in Euclidean space. We selected a user who interacted with 28 items as the subject of the case study. By analyzing items recommended by Hgformer from both head and tail positions, we aim to understand why Hgformer recommended these items while other models did not.  It can be observed in Fig~\ref{head_tail} that both the Euclidean space models and hyperbolic space models exhibit similar performance for head items, with their rankings being relatively close. However, in the case of tail items, we notice that Euclidean space models tend to rank these items lower, while hyperbolic space models can more effectively identify specific tail items that User\_4 prefers. Nonetheless, HGCF fails to recommend items that are far from User\_4 in the interaction graph (i.e., items with hops greater than 5). In contrast, Hgformer introduces the Hyperbolic self-attention mechanism, enabling the model to identify distant but relevant items effectively.

%% file: sections/relatedworks.tex
\section{Related Works}\label{sec-related}
\noindent\textbf{Hyperbolic Neural Networks.}
Due to the negative curvature characteristics of hyperbolic manifolds, they have advantages over Euclidean space when representing hierarchical structures, tree-like structures, and data with long-tail distributions \cite{hie}. Fundamental operations, such as a hyperbolic linear layer, hyperbolic activation function, and hyperbolic graph convolution, are defined in pioneering research \cite{hgcn,hnn}. However, these operations must rely on mapping the embeddings from the hyperbolic manifold back to the Euclidean space before performing operations such as self-attention, linear transformation, and activation functions. FHNN(Fully Hyperbolic Neural Network) \cite{chen2021fully} defined Lorenz linear transformation and made it possible to operate fully on the hyperbolic manifold. In order to operate graph aggregation fully on the hyperbolic manifold, H2HGCN \cite{dai2021hyperbolic} applied Einstein mid-point and LGCN \cite{zhang2021lorentzian} applied Lorentzian aggregation. However, these methods require extra network parameters, which is unsuitable in the scenario of CF. In this work, we defined a lighter hyperbolic graph convolution, called LHGCN that can operate fully on a hyperbolic manifold without model parameters.

\noindent\textbf{Graph Neural Networks for CF.}
Graph Neural Networks have been widely applied in CF in recent years \cite{ngcf,lightgcn,hgcf}. NGCF \cite{ngcf}, employs multilayer graph convolutions to model high-order connectivity and intricate user-item interactions, significantly boosting the model's expressive power. Following NGCF, 
LightGCN \cite{lightgcn} simplifies traditional GCN models by removing non-linear activation functions and self-loops, improving both efficiency and performance. However, since conventional GNNs are naturally unsuitable for modeling hierarchical and scale-free structures with complex and high-dimensional relationships, HGCF \cite{hgcf}, the firstly applies hyperbolic GCNs to CF tasks to mitigate the long-tail issues. 
Following HGCF, HICF \cite{hicf} uses the hyperbolic margin ranking to improve the effectiveness of the recommendation for both head and tail items, HRCF \cite{hrcf}, improves collaborative filtering through geometric regularization, effectively addressing the over-smoothing problem in hyperbolic graph convolutions, and HCTS \cite{yang2024hyperbolic} extended HGCF to cross-domain recommendation. However, this line of work fails to capture global information and has limitations in network expressivity.

\noindent\textbf{Graph Transformer.} Considering the powerful representation capabilities of transformers to understand the global structure of the graph and capture complex interactions of nodes within the graph, researchers have proposed the framework of the graph transformer in recent years\cite{2019graphtransformer,graphformer,difformer,nodeformer,sgformer,gformer}.
However, due to the high computational complexity of the transformer, these models can only be applied to graph classification tasks with small graphs. In recent years, researchers have been seeking methods to reduce the computation complexity of the graph transformer\cite{nodeformer,sgformer,difformer} and make graph transformers possible to deal with large-scale graphs. Considering the powerful representation capacities of the graph transformer, several searches also applied it to recommendation tasks\cite{gformer,sigformer,lightgt}. 
However, they only focus on local structure modeling and fail to capture the global structure of graphs.
Concurrently with our research, we have noted the publication of a similar work, hypformer\cite{yang2024hypformer}, which proposes a hyperbolic transformer with a linear computational complexity. While there are certain parallels between their approach and ours, there are notable differences: 1) They focus on node classification, whereas our Hgformer is specifically tailored for CF; 2) We propose LHGCN for effective local structure modeling, which is not covered in their work; 3) We provide theoretical proof for the linear approximation of our hyperbolic cross-attention mechanism. Overall, we are the first to propose a hyperbolic graph transformer in the context of collaborative filtering.

%% file: proofs/theorem1proof.tex
\noindent\textbf{Proof.}

We set $\boldsymbol{\omega} \sim \mathcal{N}(\mathbf{0}_d, \mathbf{I}_d)$ $\in \mathbb{R}^d$. For any $\mathbf{x} \in \mathbb{R}^d$ we have

\begin{equation}
    (2 \pi)^{-d / 2} \int_{\boldsymbol{\omega}} \exp \left(-\|\boldsymbol{\omega}-\mathbf{x}\|_\mathcal{E}^2 / 2\right) d \boldsymbol{\omega}=1
    \label{normal}
\end{equation}

and for any $\mathbf{x} \in \mathcal{H}^{d+1}$ : $-x_0^2+x_1^2+\cdots x_d^2=-K$

Then by definition of HSM function, we have:

\begin{equation}
    \operatorname{HSM}(\mathbf{x}, \mathbf{y})=\exp\left(\left<\mathbf{x}, \mathbf{y}\right> _{\mathcal{M}}\right)
\end{equation}

\begin{equation}
    \begin{split}
    =\exp(-x_0\cdot y_0)\exp \left(-\|\widetilde{\mathbf{x}}\|^2_{\mathcal{E}} / 2\right) \exp \left(\|\widetilde{\mathbf{x}}+\widetilde{\mathbf{y}}\|^2_{\mathcal{E}} / 2\right) \cdot\\
       \exp \left(-\|\widetilde{\mathbf{y}}\|^2_{\mathcal{E}} / 2\right)
\end{split}
\end{equation}

multiple $1$ and utilize Eq. \ref{normal}, we get:

\begin{equation}
    \begin{split}
       =\exp\left(-(2x_0y_0+\sum_{i=1}^d x_i^2+\sum_{i=1}^d y_i^2)/2\right)(2 \pi)^{-d / 2}\cdot \\
\exp \left(\|\widetilde{\mathbf{x}}+\widetilde{\mathbf{y}}\|^2_{\mathcal{E}} / 2\right) \int_{\boldsymbol{\omega}}  \exp \left(-\|\boldsymbol{\omega}-(\widetilde{\mathbf{x}}+\widetilde{\mathbf{y}})\|^2_{\mathcal{E}} / 2\right) d \boldsymbol{\omega}\\ 
    \end{split}
\end{equation}

expand it inside:

\begin{equation}
    \begin{split}
      =\exp\left(\frac{-(2x_0y_0+x_0^2-K+y_0^2-K)}{2}\right)(2 \pi)^{-d / 2} \\
      \int_{\boldsymbol{\omega}}  \exp (-\|\boldsymbol{\omega}\|^2_{\mathcal{E}} / 2+\boldsymbol{\omega}^{\top}(\widetilde{\mathbf{x}}+
      \widetilde{\mathbf{y}})-\|\widetilde{\mathbf{x}}+\widetilde{\mathbf{y}}\|^2_{\mathcal{E}} / 2+
  \\
    \|\widetilde{\mathbf{x}}+
      \widetilde{\mathbf{y}}\|^2_{\mathcal{E}} / 2) d \boldsymbol{\omega}
    \end{split}
\end{equation}

simplify the expression:

\begin{equation}
    \begin{split}
        =\exp\left(\frac{-(x_0+ y_0)^2+2K}{2} \right)(2 \pi)^{-d / 2} \\
        \int_{\boldsymbol{\omega}}  \exp \left(-\|\boldsymbol{\omega}\|^2_{\mathcal{E}} / 2+\boldsymbol{\omega}^{\top}(\widetilde{\mathbf{x}}+\widetilde{\mathbf{y}})\right) d \boldsymbol{\omega}
    \end{split}
\end{equation}

show this integration in a form similar to Eq. \ref{normal}

\begin{equation}
    \begin{split}
        =\exp\left(\frac{-(x_0+ y_0)^2+2K}{2} \right)(2 \pi)^{-d / 2}\\
        \int_{\boldsymbol{\omega}} \exp \left(-\|\boldsymbol{\omega}\|^2_{\mathcal{E}} / 2\right) \cdot \exp \left(\boldsymbol{\omega}^{\top} \widetilde{\mathbf{x}}\right) \cdot \exp \left(\boldsymbol{\omega}^{\top} \widetilde{\mathbf{y}}\right) d \boldsymbol{\omega}
    \end{split}
\end{equation}

So it is the expectation of $\left[\exp (\boldsymbol{\omega}^{\top}\left(\widetilde{\mathbf{x}}+\widetilde{\mathbf{y}}\right)\right]$:

\begin{equation}
        =\exp\left(\frac{-(x_0+ y_0)^2+2K}{2} \right)\mathbb{E}_{\boldsymbol{\omega} \sim \mathcal{N}\left(\mathbf{0}, \mathbf{I}_d\right)}\left[\exp (\boldsymbol{\omega}^{\top}\left(\widetilde{\mathbf{x}}+\widetilde{\mathbf{y}}\right)\right]
\end{equation}

which shows that is an unbiased estimation of HSM function.
\qed

%% file: proofs/lemma2proof.tex
\noindent\textbf{Proof.}

By definition of function $\phi(\mathbf{x})$:

\begin{equation}
    \begin{split}
        \phi(\mathbf{x})^{\top} \phi(\mathbf{y})=\exp(\frac{2K-x_0^2-y_0^2}{2})\\
    \frac{1}{m}\left[\exp \left(\mathbf{\boldsymbol{\omega}}_1^{\top} (\widetilde{\mathbf{x}}+\widetilde{\mathbf{y}})\right), \cdots, \exp \left(\mathbf{\boldsymbol{\omega}}_m^{\top} (\widetilde{\mathbf{x}}+\widetilde{\mathbf{y}}\right))\right]
    \end{split}
\end{equation}

\noindent since the error between $\exp(2K-x_0^2-y_0^2/2)$ and $\exp\left(-(x_0+ y_0)^2+2K/2 \right)$ are small, and the remaining part is the sampling function of the expectation, then we can take the approximation:

\begin{equation}
      \approx\exp\left(\frac{-(x_0+ y_0)^2+2K}{2} \right)\mathbb{E}_{\boldsymbol{\omega} \sim \mathcal{N}\left(\mathbf{0}_d, \mathbf{I}_d\right)}\left[\exp (\boldsymbol{\omega}^{\top}\left(\widetilde{\mathbf{x}}+\widetilde{\mathbf{y}}\right)\right]
\end{equation}
which equals to $\widetilde{\kappa}(\mathbf{x}\mathbf{y})$, and by Eq. \ref{unbias}, it also equals to $\operatorname{HSM}(\mathbf{x}, \mathbf{y})$, then complete the proof.

\qed

%% file: proofs/theorem2proof.tex
\noindent\textbf{Proof.}

First, by definition of error function:

\begin{equation}
    \Delta(\mathbf{x}, \mathbf{y})=\left|\exp \left(\langle \mathbf{x}, \mathbf{y}\rangle_\mathcal{M}\right)-\phi(\mathbf{x})^{\top} \phi(\mathbf{y})\right|
\end{equation}

take an expansion:

\begin{equation}
    \begin{split}
       \Delta(\mathbf{x}, \mathbf{y}) =\left|\exp \left(\frac{2 K-\left(x_0+y_0\right)^2}{2}\right) \exp \left(\widetilde{\mathbf{x}}^{\top} \widetilde{\mathbf{y}}\right)\right. \\
\left. 
        \exp \left(\frac{2 K-x_0^2-y_0^2}{2}\right) \frac{1}{m} \sum_{i=1}^m \boldsymbol{\omega}_i^{\top}(\widetilde{\mathbf{x}}+\widetilde{\mathbf{y}})\right| 
    \end{split}
\end{equation}

we set 

\begin{equation}
       a=\exp \left(-x_0 y_0\right)
\end{equation}

 \begin{equation}
      X=\exp \left(\frac{2 K-x_0^2-y_0^2}{2}\right) \frac{1}{m} \sum_{i=1}^m \mathbf{\boldsymbol{\omega}}^{\top}_i(\widetilde{\mathbf{x}}+\widetilde{\mathbf{y}})
 \end{equation}

then we have:

\begin{equation}
        \mathbb{E}(X)=\exp \left(\frac{2 K-x_0^2-y_0^2}{2}\right){\mathbb{E}}_{\mathbf{\boldsymbol{\omega}} \sim \mathcal{N}(\mathbf{0},\mathbf{I}_d)}\left[\exp \left(\mathbf{\boldsymbol{\omega}}^{\top}(\widetilde{\mathbf{x}}+\widetilde{\mathbf{y}})\right)\right]
\end{equation}

By Markov's inequality:

\begin{equation}
        \mathbb{P}\left(\Delta \leqslant \sqrt{\frac{\exp (3 (\delta-K))}{m \epsilon}}\right) \geqslant
         1-\frac{\left[\mathbb{E}\left((X-a \mathbb{E}(X))^2\right)\right] m \epsilon}{\exp (3 (\delta-K))} 
\end{equation}

take an expansion:

\begin{equation}
   =1-\frac{\left[\mathbb{E}\left(X^2\right)+\left(a^2-2 a\right) \mathbb{E}^2(X)\right]m \epsilon}{\exp (3 (\delta-K))} 
\end{equation}

plug in the expectation:
\begin{equation}
    \begin{split}
        =1-\bigg[\frac{1}{m} \exp (\|\widetilde{\mathbf{x}}+\widetilde{\mathbf{y}}\|^2_{\mathcal{E}}) \exp (2 \widetilde{\mathbf{x}}^{\top} \widetilde{\mathbf{y}})\\
       (1+(a^2-2 a) \exp (-\|\widetilde{x}+\widetilde{y}\|^2_{\mathcal{E}})) \bigg] m\epsilon
        /\left({\exp (3 (\delta-K))}\right)
    \end{split}
\end{equation}

Since $a=\exp \left(-x_0 y_0\right)\leqslant e^{-K} $, $K\in (0, \infty)$, we get:

\begin{equation}
    \geqslant 1-\epsilon \exp \left(\|\widetilde{\mathbf{x}}+\widetilde{\mathbf{y}}\|^2_{\mathcal{E}}+2 \widetilde{\mathbf{x}}^{\top} \widetilde{\mathbf{y}}-3 (\delta-K)\right)
\end{equation}

We assume that $\|\mathbf{x}\|^2_{\mathcal{E}}\leq\delta$,$\|\mathbf{y}\|^2_{\mathcal{E}}\leq\delta$, then 

\begin{equation}
    \|\widetilde{\mathbf{x}}\|^2_{\mathcal{E}}\leq\frac{\delta-K}{2},\|\widetilde{\mathbf{y}}\|^2_{\mathcal{E}}\leq\frac{\delta-K}{2}
\end{equation} 

and we also have:

\begin{equation}
    \|\widetilde{\mathbf{x}}+\widetilde{\mathbf{y}}\|^2_{\mathcal{E}}\leqslant 2(\delta-K), \widetilde{\mathbf{x}}^{\top} \widetilde{\mathbf{y}}\leqslant \frac{\delta-K}{2}, 
\end{equation}

then the last inequality holds:

\begin{equation}
   1-\epsilon \exp \left(\|\widetilde{\mathbf{x}}+\widetilde{\mathbf{y}}\|^2_{\mathcal{E}}+2 \widetilde{\mathbf{x}}^{\top} \widetilde{\mathbf{y}}-3 (\delta-K)\right) \geqslant 1-\epsilon
\end{equation}

Finally, we prove the upper bound of the error with probability :

\begin{equation}
    \mathbb{P}\left(\Delta \leqslant \sqrt{\frac{\exp (3 (\delta-K))}{m \epsilon}}\right)\leq 1 - \epsilon
\end{equation}

which complete the proof.
 
\qed